\documentstyle[aps,preprint,epsf]{revtex}
\newcommand{\beq}{\begin{equation}}
\newcommand{\eeq}{\end{equation}}
\begin{document}
\draft
\tightenlines

\title{ Relationships between the superconducting gap, pseudogap and
transition temperature in high-T$_{\bf c}$ superconductors}

\author{M.Ya. Amusia$^{a,b}$,
V.R. Shaginyan$^{a,c}$ \footnote{E--mail: vrshag@thd.pnpi.spb.ru}}
\address{$^{a\,}$The
Racah Institute of Physics, the Hebrew University, Jerusalem 91904,
Israel;\\ $^{b\,}$ A.F. Ioffe Physical-Technical Institute, Russian
Academy of Sciences, 194021 St. Petersburg, Russia;\\ $^{c\,}$
Petersburg Nuclear Physics Institute, Russian Academy of Sciences,
Gatchina, 188350, Russia} \maketitle

\begin{abstract}
The crossover from superconducting gap to pseudogap is
considered. We show that the superconductivity is destroyed at
the temperature $T_c$, with the superconducting gap being smoothly
transformed into the pseudogap. Relations, which are of general
interest, between the maximum values of the gap $\Delta_1$, $T_c$ and
$T^*$ are established. We presents arguments that $T^*$ as a function
of the doping level $x$ is approximately a straight line that crosses
the abscissa at some point $x_{FC}$. In the vicinity of this point
$T^*$ coincides with $T_c$, and at $x=x_{FC}$ the fermion condensation
quantum phase transition takes place.  \end{abstract}


One of the most challenging problems of modern physics is the problem
of systems with a big coupling constant. This problem is of crucial
importance particularly in the quantum field theory, making even the
quantum electrodynamics to be not a self-consistent theory \cite{rf}.
The same problem persists in many-body physics. The Landau theory of
normal Fermi liquids has offered a solution of this problem by
introducing parameters which characterize the effective
interaction into the theory \cite{lan}.  Obviously, that the
applicability of a theory that is described by big coupling
constant is closely related to revealing its limitations. Usually, it
is assumed that the breakdown of the Landau theory is defined by the
Pomeranchuk stability conditions and occur when the Landau amplitudes
being negative reach its critical value.  Note that the new phase at
which the stability conditions are restored can in principle be again
described by the Landau theory. It was demonstrated rather
recently \cite{ks} that the Pomeranchuk conditions are covering
not all possible limitations: one is overlooked,
being connected with the situation when, at  temperature $T=0$, the
effective mass can become infinitely big. As it was shown for model
cases, it can happen leading to profound consequences.
Indeed, it has been demonstrated that such a
situation can take place provided the corresponding amplitude, or
effective coupling constant, being positive reaches the critical
value, producing a completely new class of strongly correlated Fermi
liquids with the Fermion condensate (FC) \cite{ks,vol} which is
separated from a normal Fermi liquid by the fermion condensation
quantum phase transition (FCQPT) \cite{ms}. However, this state can
appear only when we are dealing with the strong coupling limit where
an absolutely reliable answer cannot be given, based on pure
theoretical first principle ground. Therefore, the only way to check
the mentioned above results is to consider experimental facts which
can point to the existence of such a state.  We assume that these
facts can be find in the field of high-$T_c$ superconductors.

There is a number of experimental facts directly pointing to the
existence of a pseudogap (PG) in the electronic
excitation spectrum of high-temperature
superconductors at temperatures above the critical temperature
$T_c$ (see e.g. \cite{tim} for a recent review). This phenomenon may
be attributed to the presence of Cooper pairs at
temperatures $T\leq T^*$, while the long range phase coherence leading
to the superconductivity manifests itself only at temperatures $T\leq
T_c$ \cite{los}.  The other scenario suggests that the pseudogap has
its origin in other then superconductivity phenomena
\cite{zh}. Recent studies, based on the intrinsic tunneling
spectroscopy, have revealed that the superconducting gap (SG) does
vanish, while PG does not even change, at $T=T_c$ \cite{ky}.
It is claimed that all this speaks in favor of different origin of
two coexisting phenomena and against the precursor-superconductivity
scenario of PG \cite{ky}. On the other hand, it was observed that the
pseudogap phase is destroyed at any temperature by applying
sufficiently strong magnetic fields \cite{shi}, while a detailed
examination of this experimental fact have shown that such a behavior
is completely compatible with a superconducting origin of the
pseudogap \cite{pie}. Moreover, detailed
examination of the tunneling spectra over wide doping range showed
that PG at various doping levels of Bi-based high temperature
superconductors is predominantly of superconducting origin and
smoothly evolve from SG at $T\geq T_c$ \cite{mzo,kug,fesa}.  At
present there is no consensus concerning the key question on whether
PG has its origin in superconductivity.  Cuprates are very
complex materials and it is evident  that PG may be caused by
various reasons. Whatever the scenario of PG
is taken, the accepted approach is to deal with a special case of
PG and to explain the observed relations between the maximum value of
the gap $\Delta_1$ at $T=0$ and temperatures $T_c$ and $T^*$.
Observed experimentally \cite{kug}, the very high ratios
$2\Delta_1/T_c\simeq 28$ and $T^*/T_c\simeq 7$, are hard to explain
within the precursor-superconductivity scenario, where, for instance,
the ratio $T^*/T_c$  is not more then 1.2 \cite{lam}. By ruling out
the possibility of the precursor-superconductivity scenario, one
faces another problem: how to explain a simple relationship between
$\Delta_1$ and $T^*$ that was found for various cuprates:
$2\Delta_1/T^*\simeq 4$ \cite{kug}. Combination of these striking
experimental data present a challenging problem in the physics of
high temperature superconductivity.

In this Letter we propose a solution of this problem
by using unorthodox model based on the combination of
FCQPT \cite{ms} and the conventional theory of the
superconductivity. We consider the crossover from the superconducting
gap to the pseudogap, thus establishing relationships between the
maximum value of the gap $\Delta_1$ and temperatures $T_c$ and $T^*$.
We show that $T^*$ as a function of the doping level $x$ can be
approximated by a straight line crossing the abscissa at some point
$x_{FC}$.  The function $T^*(x)$ and $T_c(x)$ coincides in the
vicinity of the point $x=x_{FC}$, $x_{FC}$ being the point at which
FCQPT takes place \cite{ms}.

For the reader's convenience,
we have to give here a summary of the most important
points of the model \cite{ms,ams} and
start with a brief consideration of general properties of
two-dimensional electron liquid in the superconducting state, when
the system has undergone FCQPT. At $T=0$, the ground
state energy $E_{gs}[\kappa({\bf p}),n({\bf p})]$ is a
functional of the order parameter of the superconducting state
$\kappa({\bf p})$ and of the quasiparticle occupation numbers $n({\bf
p})$ and is determined by the known equation of the weak-coupling
theory of superconductivity (see e.g. \cite{til})
\beq E_{gs}=E[n({\bf p})]
+\int \lambda_0V({\bf p}_1,{\bf
p}_2)\kappa({\bf p}_1) \kappa^*({\bf p}_2) \frac{d{\bf p}_1d{\bf
p}_2}{(2\pi)^4}.\eeq Here  $E[n({\bf p})]$ is the ground-state energy
of normal Fermi liquid, $n({\bf p})=v^2({\bf
p})$ and $\kappa({\bf p})=v({\bf p})\sqrt{1-v^2({\bf p})}$.
It is assumed that the pairing interaction
$\lambda_0V({\bf p}_1,{\bf p}_2)$ is weak. Minimizing $E_{gs}$ with
respect to $\kappa({\bf p})$
we obtain the equation connecting the single-particle energy
$\varepsilon({\bf p})$ to $\Delta({\bf p},0)$,
\beq \varepsilon({\bf p})-\mu=\Delta({\bf p},0)
\frac{1-2v^2({\bf p})} {2\kappa({\bf p})}.\eeq
The single-particle energy $\varepsilon({\bf p})$ is
determined by the Landau equation,
$\varepsilon({\bf p})=\delta
E[n({\bf p})]/\delta n({\bf p})$ \cite{lan}, and
$\mu$ is the chemical potential. The equation for
superconducting gap $\Delta({\bf p},T)$ takes form
\beq \Delta({\bf p},0)
=-\int\lambda_0V({\bf p},{\bf p}_1)\kappa({\bf p}_1)
\frac{d{\bf p}_1}{4\pi^2}
=-\frac{1}{2}\int\lambda_0
V({\bf p},{\bf p}_1) \frac{\Delta({\bf p}_1,0)}
{E({\bf p}_1,0)}
\frac{d{\bf p}_1}{4\pi^2}.\eeq
Where $E({\bf p},T)=\sqrt{(\varepsilon({\bf
p})-\mu)^2+\Delta^2({\bf p},T)}.$
If $\lambda_0\to 0$, then, one has
$\Delta({\bf p,0})\to 0$, and Eq. (2) reduces to that proposed
in \cite{ks} \beq \varepsilon({\bf p})-\mu=0,\: {\mathrm {if}}\,\,\,
\kappa({\bf p})\neq 0,\,\, (0<n({\bf p})<1);
\,\,\: p_i\leq p\leq p_f\in L_{FC}.\eeq At $T=0$, Eq. (4)
defines a new state of Fermi liquid with FC
for which the modulus of the order parameter $|\kappa({\bf p})|$ has
finite values in $L_{FC}$ range of momenta $p_i\leq p\leq p_f$
occupied by FC, and $\Delta_1\to 0$ in $L_{FC}$
\cite{ks,vol,ms}. Such a state can be considered as
superconducting, with infinitely small value of $\Delta_1$ so that
the entropy of this state is equal to zero. This
state, created by the quantum phase transition, disappears at
$T>0$ \cite{ms}. It follows from Eq. (4) that the system brakes into
two quasiparticle subsystems: the first one in $L_{FC}$
range is occupied by the quasiparticles with the effective mass
$M^*_{FC}\to \infty$, while the second by
quasiparticles with finite mass $M^*_L$ and momenta $p<p_i$. If
$\lambda_0\neq0$, $\Delta_1$ becomes finite, leading to finite value
of the effective mass $M^*_{FC}$ in $L_{FC}$, which can be obtained
from Eq. (2) \cite{ms} \beq M^*_{FC}
\simeq p_F\frac{p_f-p_i}{2\Delta_1},\eeq
while the effective mass $M^*_L$ is disturbed weakly.
Here $p_F$ is the Fermi momentum.
It follows from Eq. (5) that the quasiparticle dispersion can be
presented by two straight lines characterized by the
effective masses $M^*_{FC}$ and $M^*_L$ respectively. These lines
intersect near the binding energy $E_0$ of electrons which defines
an intrinsic energy scale of the system:  \beq E_0=\varepsilon({\bf
p}_f)-\varepsilon({\bf p}_i)
\simeq\frac{(p_f-p_i)p_F}{M^*_{FC}}\simeq 2\Delta_1.\eeq
In fact, as it is seen from Eq. (5) and (6) even at $T=0$, FC, being
absorbed by the superconducting phase transition,
never exhibits its feature which is the dispersionless
plateau associated with $M_{FC}\to \infty$. As a result, FC
forms a Fermi liquid characterized by the two finite effective masses
and by the intrinsic energy scale $E_0$ \cite{ms}.

FCQPT appears in a many-electron systems at relatively low
density, when the effective interaction becomes sufficiently
large \cite{ksz}. Calculations based on simple models show that FC
can exist as a stable state separated from the normal Fermi liquid
by the phase transition \cite{ks}. In ordinary electron liquid this
interaction is directly proportional to the dimensionless parameter
$r_s\sim 1/p_Fa_B$, where $a_B$ is the Bohr radius. If $p_i\to
p_F\to p_f$, Eq. (4) determines the point $r=r_{FC}$ at which FCQPT
takes place, and $(p_f-p_i)/p_F\sim (r_s-r_{FC})/r_{FC}$ \cite{ks}.
FCQPT precedes the formation of charge-density waves or
stripes, which take place at some value $r_s=r_{cdw}$ with
$r_{FC}<r_{cdw}$, while the Wigner crystallization takes place even
at larger values of $r_s$ and leads to insulator \cite{ksz}. On the
other hand, there are charge-density waves, or stripes,
in underdoped copper oxides and finally at small doping levels one
has insulators \cite{grun}. Then, in underdoped copper oxides, the
line-shape of single-particle excitations strongly deviates from
that of normal Fermi liquid, see, e.g.  \cite{tim}. While, in the
highly overdoped regime slight deviations from the normal Fermi
liquid are observed \cite{val1}.  Moreover, recent studies of
photoemission spectra discovered an energy scale in the spectrum of
low-energy electrons in cuprates, which manifests itself as a kink in
the single-particle spectra \cite{vall,blk}. The spectra in the
energy range (-200---0) meV can be described by two straight lines
intersecting at the binding energy $E_0\sim(50-70)$ meV \cite{blk}.
All these peculiar properties are naturally explained within a model
proposed in \cite{ms,ams}. In our model, the doping level $x$
is regarded as the density of holes per unit cell.
If $a$ is the separation between near
neighbor Cu ions, then the area of unit cell is
proportional to $a^2$, and $x/a^2$ is proportional to the hole
density. We assume that
$x_{FC}$ corresponds to the highly overdoped
regime at which slight deviations from the normal Fermi liquid take
place and introduce the effective coupling constant $g_{eff}\sim
(x-x_{FC})/x_{FC}$.  According to the model, the doping level $x$ in
cuprates is related to $r_s$ and to $(p_f-p_i)$ in the following way
\cite{ms}:  \beq g_{eff}\sim \frac{(x_{FC}-x)}{x_{FC}}\sim
\frac{r_s-r_{FC}}{r_{FC}} \sim \frac{p_f-p_i}{p_F}.\eeq Here the
value of $x=x_{FC}$ corresponding to $r_{FC}$ defines the point at
which FCQPT takes place. We employ this model and Eq.  (7) to
consider relationships between the gap, pseudogap and $T_c$.

To solve Eq. (2) analytically, we
take the Bardeen-Cooper-Schrieffer (BCS) approximation for the
interaction \cite{bcs}:  $\lambda_0V({\bf p},{\bf p}_1)=-\lambda_0$ if
$|\varepsilon({\bf p})-\mu|\leq \omega_D$, and zero outside this
domain, with $\omega_D$ being the characteristic phonon energy. Such
an interaction gives good description of the maximum gap
$\Delta_1$ in the case of the d-wave superconductivity, because the
different regions with the maximum absolute value of $\Delta_1$ and
the maximal density of states can be considered as disconnected
\cite{abr}. Therefore, the gap in this region is formed by attractive
phonon interactions which depend weakly on the momenta inside these
regions. The gap becomes $p$-independent, $\Delta({\bf
p},T)=\Delta_1(T)$, $E({\bf p},T)=E(T)$ respectively, and Eq. (2)
takes the form \beq 1=\lambda_0 N_{FC}\int_0^{E_0/2}\frac{d\xi}
{E(0)}+\lambda_0 N_{L}\int_{E_0/2}^{\omega_D}\frac{d\xi}
{E(0)}.\eeq
Here $\xi\equiv\varepsilon({\bf p})-\mu$ and $N_{FC}$ is the density
of states in $L_{FC}$, or $E_0$, range. As
it follows from Eq. (5), $N_{FC}=(p_f-p_F)p_F/2\pi\Delta_1$.
The density of states $N_{L}$ in the range
$(\omega_D-E_0/2)$ has the standard form $N_{L}=M^*_{L}/2\pi$.
If $E_0\to 0$, Eq. (8) reduces to the BCS equation.
Assuming that $E_0\leq2\omega_D$ and treating the
second integral in the right hand side of Eq. (8) as a small
perturbation of the solutions, we obtain \cite{ams}
\beq\Delta_1\simeq
2\beta\varepsilon_F\frac{p_f-p_F}{p_F}
\ln(1+\sqrt{2})\left(1+\beta
\ln\frac{2\omega_D}{E_0}\right)
\propto\varepsilon_F(x-x_{FC}).\eeq
with the Fermi energy $\varepsilon_F=p_F^2/2M^*_L$, and
dimensionless coupling constant $\beta=\lambda_0 M^*_L/2\pi$.
Here we use Eq. (7) to relate $(p_f-p_F)/p_F$ with the doping level $x$.
Taking the usual values of the dimensionless coupling constant
$\beta\simeq 0.3$, and $(p_f-p_F)/p_F\simeq (x_{FC}-x)\simeq 0.2$, we
get from Eq. (9) very large value for $\Delta_1\sim
0.1\varepsilon_F$, while in normal superconducting metals one has
$\Delta_1\sim 10^{-3}\varepsilon_F$.

At finite temperatures, $T^*\leq T$, where $T^*$ is the minimum
temperature at which $\Delta_1(T^*)=0$, Eqs.  (5) and (6) are replaced
by the following one \cite{ms}
\beq
M^*_{FC}\simeq p_F\frac{p_f-p_i}{4T};\,\,
E_0\simeq 4T.\eeq
To obtain the relationship between $T^*$ and $\Delta_1$,
we use the generalization of Eq. (8)
to finite temperatures. Then we put $T\to T^*$, that is
$\Delta_1(T\to T^*)\to 0$,  and have
\beq 1=\lambda_0 N_{FC}\int_0^{E_0/2}
\frac{d\xi}
{\xi}\tanh\frac{\xi}{2T}
+ \lambda_0 N_{L}\int_{E_0/2}^{\omega_D}
\frac{d\xi}
{\xi}\tanh\frac{\xi}{2T}.\eeq
Here, $M^*_{FC}$ and $E_0$ are given by
Eq. (10). We take into account Eq. (9) and
obtain from Eq. (11) \beq2\Delta_1/T^*\simeq 4.\eeq
Consider now the critical temperature $T_c$. One can define $T_c$ as
the temperature when $\Delta_1(T_c)\equiv 0$, then, obviously
$T_c=T^*$. Also, $T_c$ may be defined as a temperature at which
the superconductivity vanishes. Thus, we have two different
definitions, which in the case of the d-wave superconductivity can
lead to different $T_c$ values because of the special
behavior of the gap at the line of nodes \cite{sh}. Provided the
pairing interaction $\lambda_0V({\bf p}_1,{\bf p}_2)$ is the
combination of attractive interaction and repulsive one,
$V=V_{sr}+V_{lr}$, the d-wave superconductivity can take place, see
e.g. \cite{abr}.  We assume that the long-range component $V_{lr}$ of
the pairing interaction  is repulsive and has such radius $p_{lr}$ in
the momentum space that $p_F/p_{lr}\leq 1$. The short-range
component $V_{sr}\sim \lambda_0$ is relatively large and attractive,
with its radius $p_F/p_{sr}\gg 1$. Now the gap depends on the angle
$\theta$, which we reckon from the line of nodes of the gap, that is
$\Delta(p,\theta=0,T)=0$. At finite temperatures, Eq. (3) takes form
\cite{til} \beq \Delta({\bf p},T) =-\frac{1}{2}\int\lambda_0 V({\bf
p},{\bf p}_1) \frac{\Delta({\bf p}_1,T)} {E({\bf
p}_1,T)}\tanh\frac{E({\bf p}_1,T)}{2T} \frac{d{\bf p}_1}{4\pi^2}.\eeq
Hereafter, we consider solutions of Eq.
(13) in the vicinity of the line of nodes.
We transform Eq. (13) by setting $p\approx p_F$ (because the gap
$\Delta\neq 0$ only in the vicinity of the Fermi surface) and
separating the contribution $I_{lr}$ coming from $V_{lr}$, from the
contribution related to $V_{sr}$ and denoted as $I_{sr}$. At small
angles $\theta$, $I_{lr}$ can be approximated by $I_{lr}=\theta A
+\theta^3 B$, with the parameters $A$ and $B$ being independent upon
$T$, because they are defined by the integral over the regions
occupied by FC. Thus, we have \beq\Delta(p_F,\theta,T)=I_{sr}
+I_{lr}=-\int_0^{2\pi}\int
V_{sr}(\theta,p_1,\phi_1)\kappa(p_1,\phi_1)
\tanh{E(p_1,\phi_1)\over 2T}\frac{p_1dp_1d\phi_1}{4\pi^2}+\theta
A+\theta^3 B.\eeq
Now we show that at temperatures above some temperature $T_n$,
the solution of Eq. (14) has the second line of nodes at
$\theta_c(T)$ in the vicinity of the line of nodes. To show this, we
reduce Eq. (14) to an algebraic equation. We have $I_{sr}\sim
(V_{sr}\Delta_1/T)\theta$ because $\tanh(E/2T)\approx E/2T$ for $E\ll
T$ and $T\approx T_n$.
Because of the small radius of $V_{sr}$ the integration in Eq. (14)
runs over a small area located at the gap node. Dividing both parts
of Eq. (12) by $\kappa(\theta)$, we obtain \cite{sh} \beq E(\theta)
=-\left({V_0\over T}-A_1-\theta^2 B_1\right)|\theta|, \eeq where
$A_1=A/\kappa$ and $B_1=B/\kappa$ and $V_0\sim V_{sr}(0)\Delta_1$ is
a constant. Imposing the condition that Eq. (13) has the only
solution $\Delta\equiv 0$ when $V_{sr}=0$, we see that $A_1$ is
negative. The factor in the brackets on the right-hand side of Eq.
(15) changes its sign at some temperature $T_n\sim V_0/A_1$.
Because the excitation energy must be $E(\theta)>0$, the sign of
$\Delta$ must be reversed at the point $\theta=\theta_c$, with
$\Delta(\theta_c)=\Delta(0)=0$. The gap $\Delta$ has an entire new
line of nodes at $T\geq T_n$, with $\theta_c\to0$ at $T\to
T_n$, see Fig. 1. The gap $\Delta(\theta)$ is very
small in the region $[0<\theta<\theta_c]$ and can be destroyed only
within this region because of the different reasons, for instance, by
spin density waves or ferromagnetic fluctuations \cite{sh,ars}.
This extinction of the gap opens the channel for decay of the
supercurrent. We can conclude that $T_c\sim T_n$ is the temperature
at which the superconductivity vanishes, and a strong variation of
the superconductivity characteristics may be observed when $T\to
T_n$. At high levels of overdoping, that is $x\to x_{FC}$,
FC plays unimportant role, the coefficient $A_1$
is small and $T_n$ has no sense because $T_n>T^*$.
Therefore, $T_c=T^*$. As soon as the difference $(x-x_{FC})$ increases,
$T_n<T^*$, and we have $T_c<T^*$.
Note, that even though $T_c\simeq T^*$, a pseudogap-like behavior can
be observed because of the ratio $M^*_L/M^*_{FC}\ll 1$, which leads to
a strong change in the density of states at $E_0\simeq 2\Delta_1$,
see Eq. (10). We conclude that at $T_c\leq T$ when pseudogap is
small or absent, it is the scale $E_0$ that produces the
pseudogap-like structure in the quasiparticle density of state at the
energy $\simeq 2\Delta_1$. In case of the presence of
pseudogap, the strong change in the density of states is defined
principally by the pseudogap, as it follows from the above
consideration. At $T\leq T_c$, the superconducting peak
emerges in the density of states. With decreasing temperature it
moves to higher energies until it reaches its maximum value
$2\Delta_1$, merging with the kink in the density of states at $E_0$,
as it was observed in the experiments \cite{ky}.

At the underdoped levels, there are strong antiferromagnetic
correlations, and the gap can be destroyed locally even at $T<T_n$,
because the gap becomes small around its line of nodes at $\theta=0$.
Therefore, in this domain of $x$, $T_c(x)<T_n(x)$ and the function
$T_c(x)$ is no longer proportional to $\Delta_1$. As it follows from
Eq. (9), one has $\Delta_1\sim (p_f-p_i)/p_F\sim(x_{FC}-x)$, and
$\Delta_1$ continues to grow with the doping decreasing. A detailed
analysis of this situation will be published elsewhere.
At $T^*>T>T_{c}$, the gap occupies only a part of the Fermi surface,
and SG smoothly transforms into the PG.
As we have seen above, the ratio $2\Delta_1/T_c\simeq
2\Delta_1/T_n$  and can reach very high values. For instance,
in the case of overdoped Bi$_2$Sr$_2$CaCu$_2$Q$_{6+\delta}$, where the
superconductivity and the pseudogap are considered to be of the
common origin, $2\Delta_1/T_c$ is about 28, while the ratio
$2\Delta_1/T^*\simeq 4$; this ratio is also equal to 4 in the
case of various cuprates \cite{kug}. From Eq. (9) it follows that
$\Delta_1$ is directly proportional to $(x_{FC}-x)\sim (p_f-p_i)/p_F$,
and one finds from Eq. (12) that the function $T^*(x)$ presents
approximately a straight line crossing the abscissa at the point
$x\simeq x_{FC}$, while in the vicinity of this point $T^*(x)$
coincides with $T_c(x)$. Note, that at $(x-x_{FC})\to 0$ the scale
$E_0\to 0$ as well and the main contribution to the gap comes from
the second term on the right-hand side of Eq. (8). This
contribution, being obviously small, changes slightly the line around
$x_{FC}$ point. Strong deviations from the straight line behavior can
also take place in the case of strongly underdoped samples
because of vanishing the superconductivity and
approaching the insulator regime.

This work was supported in
part by the Russian Foundation for Basic Research, No 01-02-17189.

\begin{figure}[t]
\centerline{\epsfxsize=16cm
\epsfbox{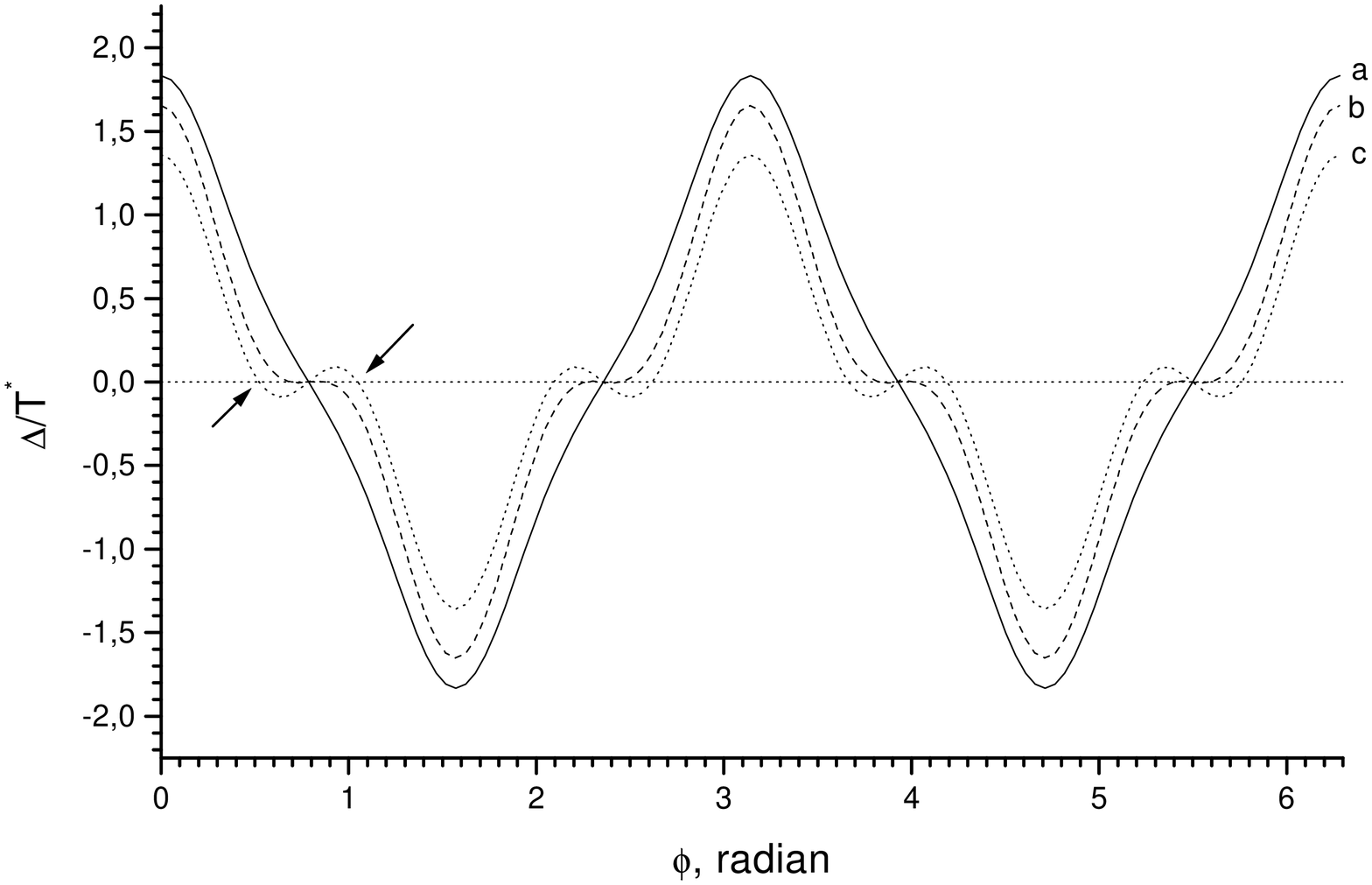}}
\caption{The gap $\Delta$
as a function of $\phi=\theta+\pi/4$ calculated at three different
temperatures expressed in terms of $T_{n}\simeq T_c$, while
$\Delta$ is presented in terms of $T^*$. Curve (a), solid line, shows
the gap calculated at temperature $0.9 T_{n}$. In curve (b),
dashed line, the gap is given at $T_{n}$.
$\Delta(\phi)$ at $1.2 T_{n}$ is
shown by curve (c), dotted line. The arrows indicate the two nodes
at the Fermi surface emerged at $T_{n}$.} \end{figure}
\end{document}